\documentclass[review]{elsarticle}

\usepackage{lineno,hyperref}
\usepackage{balance}
\usepackage{subfigure}
\usepackage{makecell}
\usepackage{amsmath,amssymb,amsfonts,bm}
\usepackage{multirow}
\usepackage{tablefootnote}
\usepackage{color}
\usepackage{algorithm}
\usepackage{algorithmic}
\usepackage{multicol}
\usepackage{booktabs}
\usepackage{appendix}

\journal{Journal of \LaTeX\ Templates}

\begin{document}

\begin{frontmatter}

\title{Modeling Multi-aspect Preferences and Intents for Multi-behavioral Sequential Recommendation}




\author[mymainaddress,mysecondaryaddress]{Haobing Liu}
\ead{haobingliu@ouc.edu.cn}

\author[mysecondaryaddress]{Jianyu Ding}
\ead{dingjianyu@sjtu.edu.cn}

\author[mysecondaryaddress]{Yanmin Zhu\corref{mycorrespondingauthor}}
\cortext[mycorrespondingauthor]{Corresponding author}
\ead{yzhu@sjtu.edu.cn}

\author[mysecondaryaddress]{Feilong Tang}
\ead{tang-fl@cs.sjtu.edu.cn}

\author[mysecondaryaddress]{Jiadi Yu}
\ead{jiadiyu@sjtu.edu.cn}

\author[mymainaddress]{Ruobing Jiang}
\ead{jrb@ouc.edu.cn}

\author[mymainaddress]{Zhongwen Guo}
\ead{guozhw@ouc.edu.cn}

\address[mymainaddress]{Department of Computer Science and Technology, Ocean University of China}
\address[mysecondaryaddress]{Department of Computer Science and Engineering, Shanghai Jiao Tong University}

\begin{abstract}
Multi-behavioral sequential recommendation has recently attracted increasing attention. However, existing methods suffer from two major limitations. Firstly, user preferences and intents can be described in fine-grained detail from multiple perspectives; yet, these methods fail to capture their multi-aspect nature. Secondly, user behaviors may contain noises, and most existing methods could not effectively deal with noises. In this paper, we present an attentive recurrent model with multiple projections to capture \textbf{M}ulti-\textbf{A}spect preferences and \textbf{INT}ents (MAINT in short). To extract multi-aspect preferences from target behaviors, we propose a multi-aspect projection mechanism for generating multiple preference representations from multiple aspects. To extract multi-aspect intents from multi-typed behaviors, we propose a behavior-enhanced LSTM and a multi-aspect refinement attention mechanism. The attention mechanism can filter out noises and generate multiple intent representations from different aspects. To adaptively fuse user preferences and intents, we propose a multi-aspect gated fusion mechanism. Extensive experiments conducted on real-world datasets have demonstrated the effectiveness of our model.
\end{abstract}


\end{frontmatter}

\section{Introduction}
With the rapid growth of the amount of information, recommender systems seeking to predict items that a user may have an interest in have become fundamental for helping users overcome information overload. Since user preferences may change over time, sequential recommendation has been studied, which can model sequence-related patterns (such as sequential patterns, co-occurrence patterns, and distance patterns) in user-item interactions~\cite{QuadranaCJ18}.  

Recently, various types of user interactions have increasingly been collected~\cite{gao2021learning,LiuZZXYT22,LiuZWDYT23}. For example, an e-commerce website may collect click, add-to-cart, and buy behaviors. However, most existing recommender models only consider \emph{the target type of behavior}, which directly relates to the KPI and is believed to be the strongest signal reflecting a user's preference. As such, \emph{support types of behaviors} are often neglected. This is problematic because support types of behaviors can provide vital clues about a user's intent. Both user preference (what users love) and intent (what users currently want) are critical factors that influence whether or not a user is interested in an item~\cite{wang2020toward}.

To leverage multi-typed behaviors, various existing methods have been proposed and they can be divided into two categories. The first category mainly involves extending the Matrix Factorization (MF) algorithm~\cite{koren2009matrix}, Bayesian Personalized Ranking (BPR) algorithm~\cite{rendle2009bpr}, or using Multi-Layer Perceptron (MLP) to seek novel solutions. However, we argue that this category models multi-typed behaviors from a static perspective, which neglects sequence-related patterns in user-item interactions. 
The second category considers both the multi-typed and sequential nature of behaviors, and can be further divided into two subcategories. The first subcategory regards all behaviors as one long sequence and designs methods that aim to handle it~\cite{zhou2018micro, liu2017multi}. However, this subcategory fails to capture the intrinsic patterns of each type of behavior. The second subcategory models at least two kinds of sequences to alleviate the limitations of the first subcategory~\cite{li2018learning, guo2019buying,tanjim2020attentive}. However, it still fails to consider multiple aspects of user preferences and intentions. Besides, most methods ignore that support types of behavior may be very noisy. If noises are not handled properly, they may harm the performance of models. 

\begin{figure*}[t]
\centering
    \includegraphics[width= 12cm]{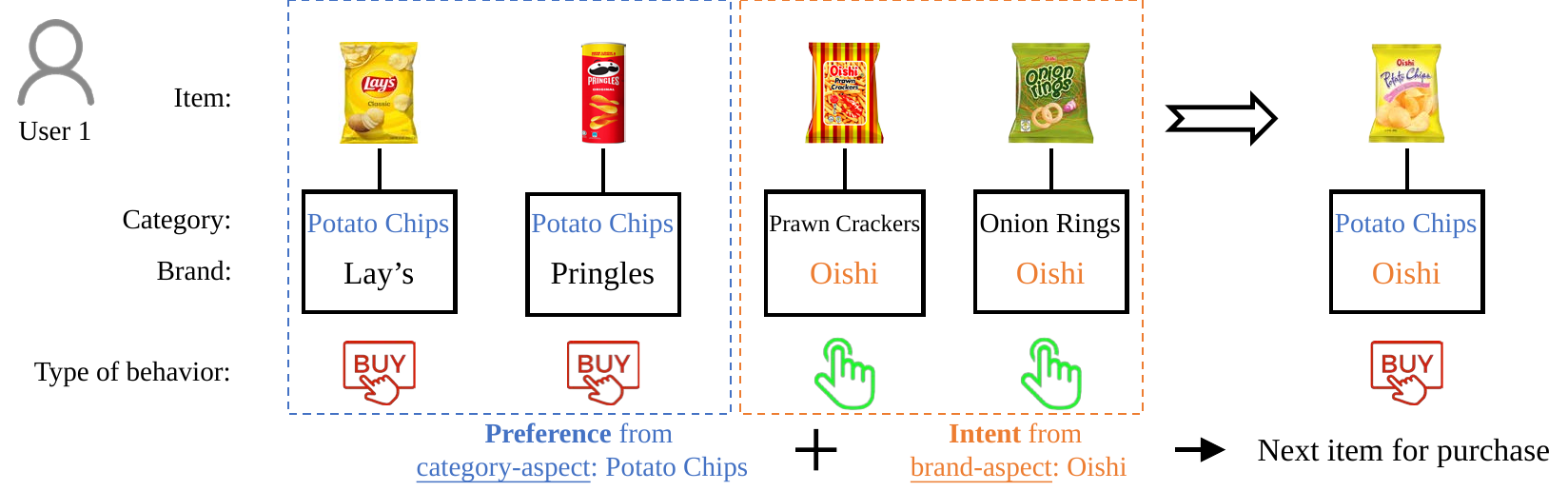}
    \caption{Recommending by integrating multi-aspect preferences and intents.}
    \label{fig:illu}
\end{figure*}

Actually, user preferences and intents are multi-aspect. In other words, user preferences and intents should be described in a fine-grained way. This can be better understood referring to Figure~\ref{fig:illu} as an example. User preferences and intents are described from two main aspects: the category-aspect and the brand-aspect. The historical buy behaviors of user 1 imply that she loves potato chips. Additionally, her click records indicate that she is shopping for Oishi products. By considering both the preferences and intents of user 1, it would be appropriate to recommend Oishi potato chips to her.

We find three key technical challenges in multi-behavioral sequential recommendation. 
First, we try to learn user intents from her multi-typed behavioral sequence. However, different behavioral specifics (e.g., behavior type, time interval) which provide a fine-grained description of one user interaction on one item indicate different strengths of the intents. For example, \emph{add-to-cart} could be a stronger signal compared with \emph{click}. Besides, the multi-typed behavioral sequence can be noisy. To support this observation, we calculate the conversion rates of different behavior types using real-world datasets. The results, shown in Table~\ref{tab:coversionRate}, reveal extremely small conversion rates of \emph{click}, \emph{collect}, and \emph{view}. This indicates that noises can be introduced when utilizing support types of behaviors to help to predict the next target behavior. Thus, a challenge lies in \emph{how to extract users' intents from multi-typed behavioral sequences while considering the behavioral specifics and the noisy nature of the sequences}. 
Second, user preferences and intents are multi-aspect. Figure~\ref{fig:illu} provides an example that illustrates how user preferences and intents are reflected in the category-aspect and the brand-aspect. While explicit aspects like category and brand help explain the multi-aspect nature of preferences and intents, these aspects are implicit in the real world. Leveraging one latent vector to represent a user's preferences/intents is not enough to capture the multi-aspect nature. Thus, the second challenge is on \emph{how to capture the multi-aspect nature of users' preferences and intents}. 
Third, preferences and intents have different degrees of impact on different users. Considering users who only made few purchases a long time ago but clicked many items recently, their intents would be more important compared with their preferences. Simple fusion methods (e.g., concatenation~\cite{li2018learning, guo2019buying}) may be inappropriate. Thus, \emph{how to fuse multi-aspect preferences and intents adaptively} is challenging.

\begin{table}[t]
    \caption{Conversion rates of different behavior types on two datasets.}
    \centering
    \small
    \begin{tabular}{lcc}
    \toprule
    \textbf{Dataset} & \textbf{Taobao} & \textbf{Retailrocket}\\ \hline              
    Conversion Rate & \begin{tabular}[c]{@{}r@{}}0.09 \ \ \ Click\\ 0.07 \ Collect\\ 0.26 \ \ \ \ Cart\end{tabular} & \begin{tabular}[c]{@{}r@{}}0.16 \ View\\ 0.76 \ \ Cart\end{tabular} \\ 
    \bottomrule
    \end{tabular}
    \label{tab:coversionRate}
\end{table}

To address the aforementioned challenges, we propose a novel attentive recurrent model with multiple projections for capturing \textbf{M}ulti-\textbf{A}spect preferences and \textbf{INT}ents (MAINT in short). 
\emph{To model user intents from noisy multi-typed behavioral sequences}, we design a behavior-enhanced LSTM and a refinement attention mechanism. By adding behavioral specifics in the gates of Long Short-Term Memory (LSTM)~\cite{hochreiter1997long}, LSTM can consider behavioral specifics while modeling an item sequence. The refinement attention mechanism regards the stable preference as a guider to filter out noises during intent extraction. 
\emph{To capture the multi-aspect nature of preferences and intents}, we propose a multi-aspect projection mechanism and leverage parallel attention mechanisms. In detail, we first project the hidden state of the LSTM modeling the target behavior sequence to multiple semantic subspaces representing multiple aspects. This results in multiple representations of hidden user preferences from different aspects. Then, we regard these preferences as guiders and perform the refinement attention mechanisms in parallel, obtaining representations of multi-aspect latent intents. 
\emph{To adaptively fuse multi-aspect preferences and intents}, we develop a multi-aspect gated fusion mechanism. The mechanism can balance the impacts of preferences and intents by considering their characteristics in each aspect.

The main contributions of this work are as follows:
\begin{itemize}
    \item We propose a novel approach for multi-behavioral sequential recommendation. To the best of our knowledge, it is the first model which can extract and fuse multi-aspect preferences and intents.
    \item To model multi-aspect preferences, we design a multi-aspect projection mechanism that generates multiple preference representations from different aspects.
    \item To model multi-aspect intents, we design a behavior-enhanced LSTM and a multi-aspect refinement attention mechanism. The LSTM captures specific while modeling an item sequence. The attention mechanism generates multiple intent representations from different aspects.
    \item To fuse multi-aspect preferences and intents, we propose a multi-aspect gated fusion mechanism. In each aspect, there is a gating mechanism that integrates the preferences and intents from that aspect.
\end{itemize}


\section{Related Work}\label{rw}
We classify related studies into three categories: multi-behavioral non-sequential recommendation, single-behavioral sequential recommendation, and multi-behavioral sequential recommendation as Table~\ref{tab:cat} shows. 

\begin{table}[t]
\caption{Categorization of related studies.}
\label{tab:cat}
\centering
\small
    \begin{tabular}{l|l|l}
    \hline
    \textbf{Category}                                                                                          & \textbf{Approaches}                                                                               & \textbf{References}                                                                           \\ \hline
    \multirow{2}{*}{\begin{tabular}[c]{@{}l@{}}Multi-behavioral Non-sequential \\ Recommendation\end{tabular}} & Traditional Methods                                                                               & \cite{zhao2015improving,loni2016bayesian}                                    \\ \cline{2-3} 
                                                                                                               & Deep Learning-based Methods                                                                       & \cite{gao2021learning,jin2020multi,xia2021multi,chen2021graph,ZhangMCX20,ParkK0Y20,YuFYHZD22}    \\ \hline
    \multirow{6}{*}{\begin{tabular}[c]{@{}l@{}}Single-behavioral Sequential \\ Recommendation\end{tabular}}    & Traditional Methods                                                                               & \cite{rendle2010factorizing,HeM16,HosseiniYZSKC19}                           \\ \cline{2-3} 
                                                                                                               & RNN-based Methods                                                                                 & \cite{hidasi2016session,zhu2017next,wang2019modeling,ZhaoLLXLZSZ22}                        \\ \cline{2-3} 
                                                                                                               & CNN-based Methods                                                                                 & \cite{DBLP:conf/wsdm/TangW18,DBLP:conf/wsdm/YuanKAJ019}                      \\ \cline{2-3} 
                                                                                                               & Transformer-based Methods                                                                         & \cite{kang2018self,zhang2019feature,LuoLL21,saaki2022value}                          \\ \cline{2-3} 
                                                                                                               & GNN-based Methods                                                                                 & \cite{WuT0WXT19,LiCLYH21,RaoCLSYH22}                                                    \\ \cline{2-3} 
                                                                                                               & Other Methods                                                                                     & \cite{MaKL19}                                                                \\ \hline
    \multirow{2}{*}{\begin{tabular}[c]{@{}l@{}}Multi-behavioral Sequential \\ Recommendation\end{tabular}}     & \begin{tabular}[c]{@{}l@{}}Methods Taking One Kind of \\ Sequence as Input\end{tabular}           & \cite{liu2017multi,zhou2018micro}                                            \\ \cline{2-3} 
                                                                                                               & \begin{tabular}[c]{@{}l@{}}Methods Taking At Least Two \\ Kinds of Sequence as Input\end{tabular} & \cite{li2018learning,zhou2018atrank,tanjim2020attentive,XuZY21,WuXZACZZLH22} \\ \hline
    \end{tabular}    
\end{table}

\subsection{Multi-behavioral Non-sequential Recommendation}
Multi-behavioral non-sequential recommender methods are designed to capture the heterogeneity of behaviors. According to whether utilizing deep learning structures, methods can be divided into traditional methods and deep learning-based methods.

\subsubsection{Traditional Methods}
Traditional methods are mainly based on the MF algorithm or the BPR algorithm. For example, Zhao et al.~\cite{zhao2015improving} jointly factorized multiple matrices of different behavior types with sharing item-side embeddings. Loni et al.~\cite{loni2016bayesian} proposed sampling rules considering levels of different behavior types. 

\subsubsection{Deep Learning-based Methods}
Some researchers attempt to leverage deep neural networks. For example, NMTR~\cite{gao2021learning} combines the advance of NCF~\cite{he2017neurala} with multi-task learning. For each behavior type, one specific interaction function is learned. Based on multi-typed behavioral data, a user-item bipartite graph with multiple types of edges can be constructed. Then Graph Neural Networks (GNNs) could be utilized to handle the user-item multi-behavior graph. 
Jin et al.~\cite{jin2020multi} assigned different learnable weights to different kinds of edges with graph attention networks to model the importance of the behaviors. Similarly, Xia et al.~\cite{xia2021multi} designed an aggregation mechanism which is based on the attentional neural mechanism to explicitly model the dependencies between different types. Chen et al.~\cite{chen2021graph} assigned representations to behavior types (i.e., edge types). Zhang et al.~\cite{ZhangMCX20} proposed a graph convolutional network-based method to learn user and item representations under different behavior types. Moreover, some researchers explore heterogeneous graph embedding methods~\cite{ParkK0Y20,YuFYHZD22}. We argue these methods mainly ignore sequence-related patterns of behaviors.

\subsection{Single-behavioral Sequential Recommendation}
Most sequential recommendation methods focus on handling single-behavioral sequences. They strive to model sequential dependencies existing in user interactions~\cite{fang2020deep}. We first review traditional methods. Then, we summarize Recurrent Neural Network (RNN)-based methods, Convolution Neural Network (CNN)-based methods, transformer-based methods, GNN-based methods, and other methods.

\subsubsection{Traditional Methods}
Traditional methods are mainly based on Markov Chains (MCs). For example, Rendle et al.~\cite{rendle2010factorizing} combined the MF with first-order MCs. He et al.~\cite{HeM16} further integrated an item similarity model with high-order MCs. 

Hosseini et al.~\cite{HosseiniYZSKC19} proposed a method to retrieve multi-aspect temporal similarity maps. The method aims to alleviate the sparseness issue in the user-location matrix. Additionally, they utilize the Expectation-Maximization technique to compensate for incomplete data at each temporal scale.

\subsubsection{RNN-based Methods}
RNNs have been widely adopted to model sequential data (e.g., sequences of words) and have shown efficient performance. Hidasi et al.~\cite{hidasi2016session} introduced the Gated Recurrent Unit (GRU)~\cite{chung2014empirical} to model click behaviors. Zhu et al.~\cite{zhu2017next} modified the gate structure of the vanilla LSTM cell. Wang et al.~\cite{wang2019modeling} proposed a neural network to learn multiple purposes of users. Several RNNs were used to learn different purposes and a purpose router was designed to decide which RNN should be updated with an interacted item. Check-in behaviors are spatio-temporal data. Zhao et al.~\cite{ZhaoLLXLZSZ22} extended the vanilla LSTM cell with additional spatio-temporal gates, enabling the utilization of time and distance intervals. These methods cannot ignore the heterogeneity of behaviors. 

\subsubsection{CNN-based Methods}
Besides RNNs, some researchers explore CNNs. CNNs are commonly applied to handle image data. Tang and Wang~\cite{DBLP:conf/wsdm/TangW18} regarded the embedding matrix $E\in\mathbb{R}^{L\times d}$ of $L$ previous items as an ``image''. They used a vertical horizontal convolutional layer and a horizontal convolutional layer to handle the ``image''. Yuan et al.~\cite{DBLP:conf/wsdm/YuanKAJ019} also regard an item embedding sequences as an ``image'' and explored holed CNN layers. Since CNNs are suitable to capture local features, these methods cannot well capture the dependencies between two distant interactions. 

\subsubsection{Transformer-based Methods}
Latterly, transformer~\cite{vaswani2017attention} has been introduced into recommender systems. For example, SASRec~\cite{kang2018self} is based on a self-attention network. Zhang et al.~\cite{zhang2019feature} applied parallel self-attention blocks on both item sequences and feature sequences. Luo et al.~\cite{LuoLL21} constructed spatio-temporal relation matrices and utilized a self-attention layer to capture the dependencies between non-adjacent POIs and non-contiguous visits. Saaki et al.~\cite{saaki2022value} leveraged discrete-time modeling to exploit concepts from texts and continuous-time modeling to infer infinite behavioral patterns of users. These methods model sequential dependencies with the help of the positional encoding technique. Researchers have developed various positional encoding schemes~\cite{qiu2021tois}. Choosing an improper positional encoding scheme will result in incorrect time modeling. 

\subsubsection{GNN-based Methods}
In addition, some researchers try to transform sequences into graph-structured data and utilize GNNs for sequential recommendation. For example, Wu et al.~\cite{WuT0WXT19} transformed item sequences into item-item directed graphs and leveraged a gated graph neural network to handle the graphs. 
Li et al.~\cite{LiCLYH21} constructed graph-augmented POI sequences to capture the collaborative signals from semantically correlated POIs (i.e., POIs before and after the target POI in different sequences). They proposed a position-aware attention net to model sequential dependencies. Rao et al.~\cite{RaoCLSYH22} first constructed a spatial-temporal knowledge graph and leveraged a knowledge graph embedding method to learn embeddings of entities and relations. Next, they constructed a POI transition graph based on the embeddings. Then, they incorporated the POI transition graph into RNN-based models. We argue the transformations are not one-to-one mappings, thus they may be lossy. 
Besides, these methods may not effectively capture long-term dependencies.

\subsubsection{Other Methods}
Other deep learning structures have also been adopted in the single-behavioral sequential recommendation. For example, Ma et al.~\cite{MaKL19} specifically designed a hierarchical gating network with BPR to capture both long-term and short-term user preferences. To capture the short-term user preferences, they designed a feature gating and an instance gating for hierarchically selecting features and items.

\subsection{Multi-behavioral Sequential Recommendation}
For capturing the multi-typed and sequential nature of interactions, researchers have developed multi-behavioral sequential recommender models, which can be further classified into two subcategories.

\subsubsection{Methods Taking One Kind of Sequence as Input}
The first subcategory regards all behaviors as one long sequence. 
Liu et al.~\cite{liu2017multi} combined RNN with log-bilinear model to handle the long sequence. To capture properties of multiple types of behaviors, they incorporated behavior-specific matrices. Zhou et al.~\cite{zhou2018micro} leveraged an attention-based RNN to model sequential information of the long sequence. They learned representations for each behavior type and concatenated the representations and item representations as input. We argue that these methods cannot capture the intrinsic patterns of each type of behavior well.

\subsubsection{Methods Taking At Least Two Kinds of Sequence as Input}
The second subcategory models at least two kinds of sequence. For example, Li et al.~\cite{li2018learning} proposed an LSTM-based model to learn from the preference behaviors and the session behaviors. Zhou et al.~\cite{zhou2018atrank} grouped behaviors by behavior type and leveraged a self-attention network to model the influence among different behavior types. Tanjim et al.~\cite{tanjim2020attentive} learned item similarities based on the interacted item sequence via a transformer layer and obtained the user’s intent based on her actions on a particular category via a convolution layer. Xu et al.~\cite{XuZY21} utilized multiple GRUs to learn diverse intentions. It designs three tasks to improve recommendation performance. Wu et al.~\cite{WuXZACZZLH22} adopted contrastive learning among different behavior types. We argue that these studies fail to consider the multi-aspect nature of both preferences and intents. Furthermore, most studies ignore that support types of behaviors can be noisy.

\section{Preliminaries}\label{pr}
\subsection{Problem Formulation}
\begin{table}[t]
\caption{Main notations.}
\label{tab:not}
\centering
\small
    \begin{tabular}{l|l}
    \hline
    \textbf{Notation}                           & \textbf{Description}                                                                                                                                             \\ \hline
    $x = (i, c, b, t)$                 & \begin{tabular}[c]{@{}l@{}}one behavior on item $i$ whose content information is $c$ (The behavior \\ type is $b$ and it happened at $t$.)\end{tabular} \\
    $X^u = \{x_1, x_2, \cdots , x_N\}$ & the behavior sequence belonging to user $u$                                                                                                             \\
    $p,q$              & the embedding for item $i$, item content $c$                                                                                                            \\
    $r,s$              & the embedding for behavior type $b$, time interval $\Delta t$                                                                                           \\
    $J$                & the number of implicit aspects                                                                                                                          \\
    $\tilde{h}^S_{j}$  & the projected stable preference representation in the $j$-th aspect                                                                                     \\
    $\alpha_{j,n}$     & the attention score assigned to the $n$-th item in the $j$-th aspect                                                                                  \\
    $\tilde{h}^D_{j}$  & the dynamic intent representation in the $j$-th aspect                                                                                                  \\
    $\beta_j$          & the gate scores assigned to the latent intent in the $j$-th aspect                                                                                      \\
    $\tilde{h}^H_{j}$  & the hybrid user representation in the $j$-th aspect                                                                                                     \\
    $h^F$              & the final user representation                                                                                                                           \\
    $\hat{y}^I$        & the predicted probabilities of items                                                                                                                       \\ \hline
    \end{tabular}
\end{table}

The main notations utilized throughout this paper are outlined in Table~\ref{tab:not}. While users interact with recommender systems, interactions will be recorded. Given a user $u$, interactions of $u$ can be defined as $X^u = \{x_1, x_2, \cdots , x_N\}$. The $n$-th element $x_n = (i_n, c_n, t_n, b_n)$ indicates that $u$ performs a \emph{behavior} of type $b_n$ on the item $i_n$ at the timestamp $t_n$. All users constitute a user set $U$. All items occurring in all behaviors constitute an item set $I$. All behavior types constitute a behavior type set $B$. $c_n$ is the content information (such as category information and brand information) of item $i_n$. 

In recommender systems, there exists a key type of behavior referred to as \emph{target type}. On e-commerce websites, the target type is usually \emph{buy}. For ease of illustration, we use ``buy'' or ``target'' interchangeably throughout this paper. The other types of behaviors belong to \emph{support types}. 

\noindent\textbf{Problem Formulation.} 
The problem of multi-behavioral sequential recommendation is formulated as follows:

\textbf{Input:} Users $U$, items $I$, behavior types $B$, and all users’ behavioral data $X$. 

\textbf{Output:} A predictive model aims to predict the item that each user is most likely to interact with under the target behavior type.

\subsection{Framework Overview}
\begin{figure*}[t]
\centering
    \includegraphics[width= 8cm]{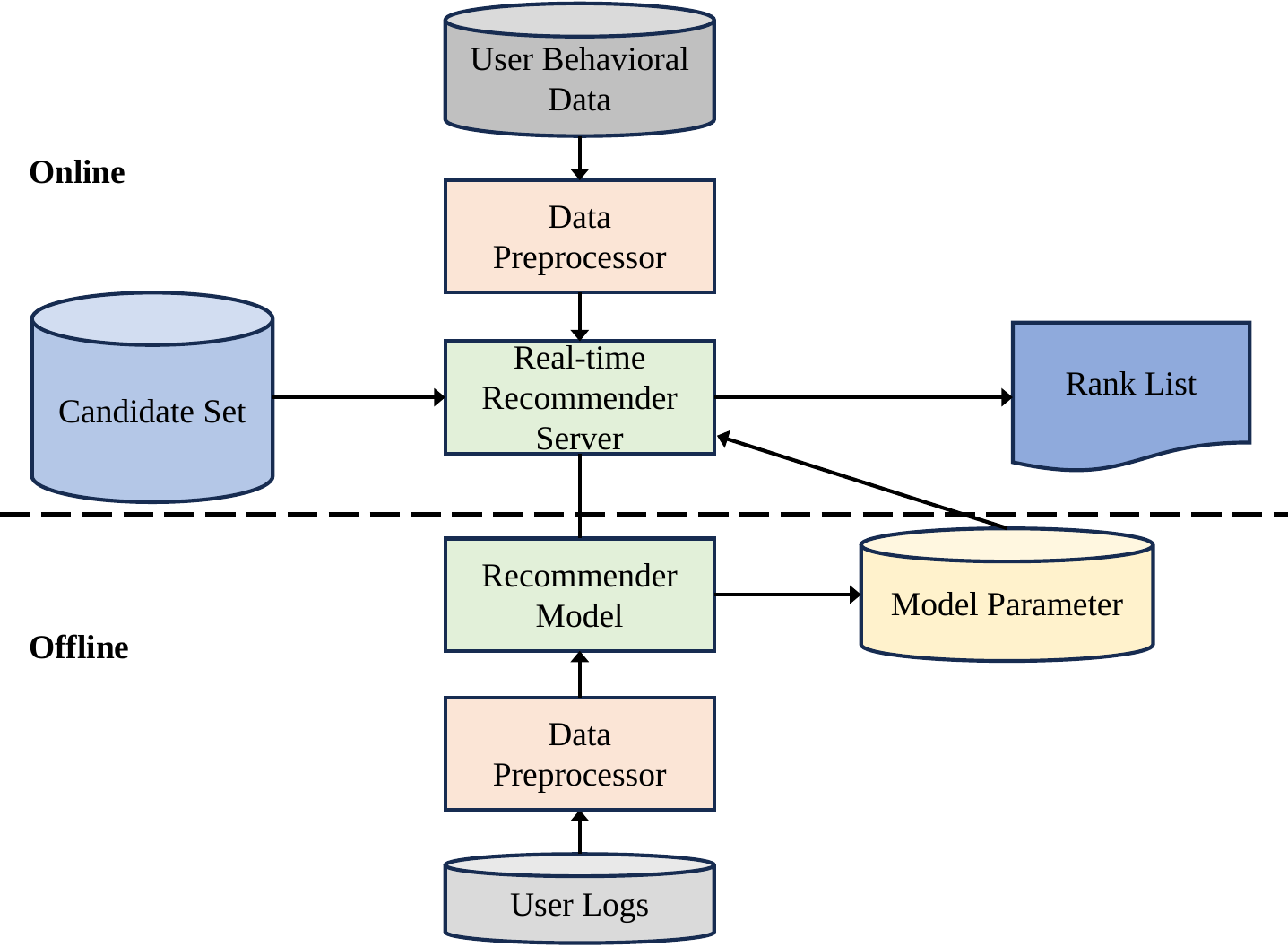}
    \caption{Framework.}
    \label{fig:framework}
\end{figure*}

Figure~\ref{fig:framework} illustrates our framework, consisting of two phases: offline and online, to recommend the top K items that each user is most likely to interact with under the target behavior type.
\begin{itemize}
    \item \textbf{Offline Phase.} The Data Preprocessor processes User Logs in accordance with the supported input format to generate inputs for the Recommender Model. The Recommender Model is trained using the training data to obtain well-trained model parameters.
    \item \textbf{Online Phase.} The Data Preprocessor processes User Behavioral Data to generate inputs for the model. The model is deployed on servers and the trained model parameters are loaded. The Real-time Recommender Server ranks candidate items based on each user's historical behavior and presents the user with a sorted list of the top-ranked items.
\end{itemize}

\section{Proposed Model}\label{mo}
\begin{figure*}[t]
\centering
    \includegraphics[width= 12cm]{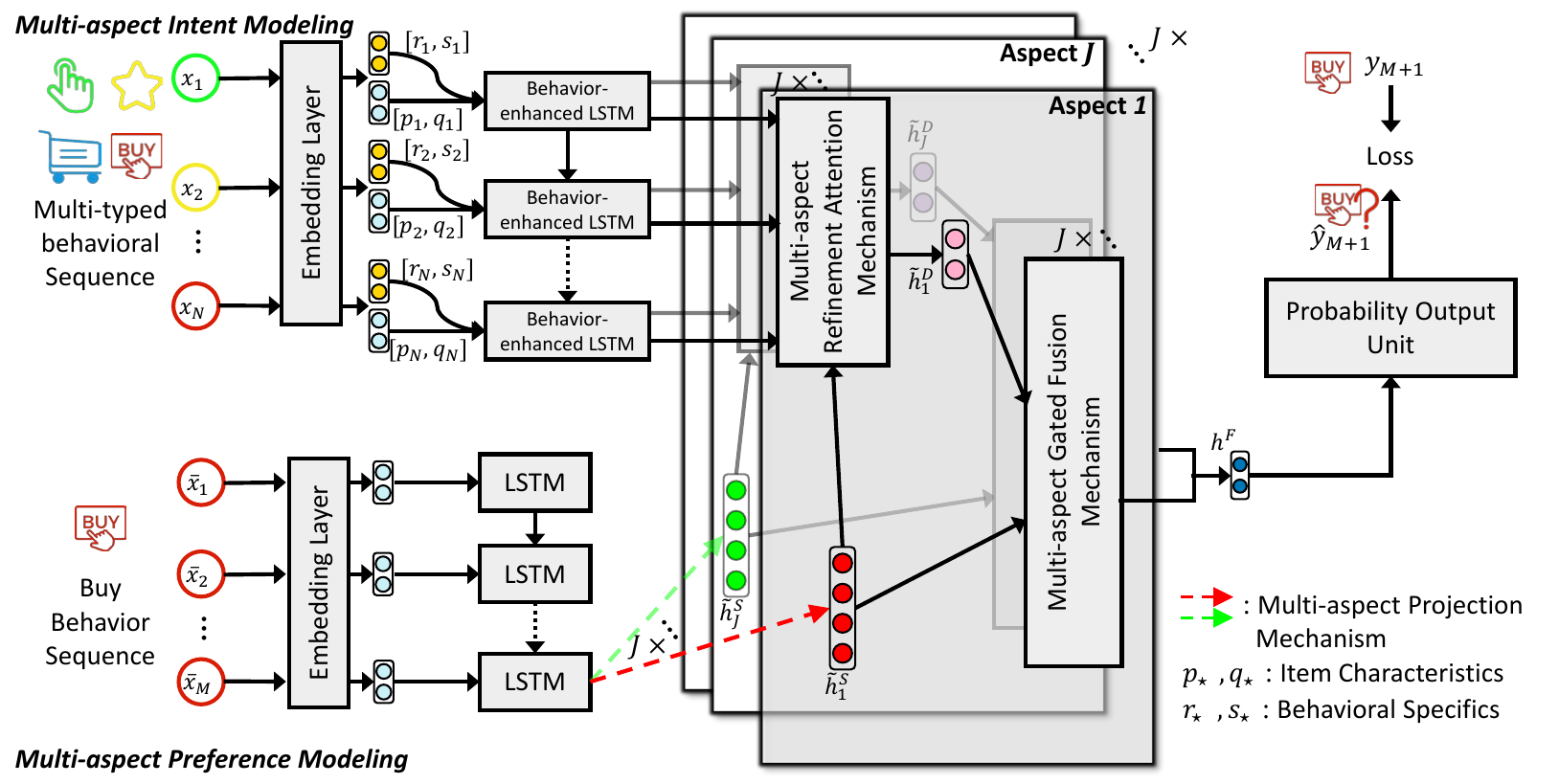}
    \caption{The overview of the MAINT structure. Figure best viewed in color.}
    \label{fig:structure}
\end{figure*}

We next present the detailed design of our model, whose structure is presented in Figure~\ref{fig:structure}. Our model has three components: 1) multi-aspect preference modeling component; 2) multi-aspect intent modeling component; 3) preference and intent fusing component. We will illustrate our model using the example from Figure~\ref{fig:illu}.

\subsection{Shared Embedding}
A shared embedding layer is utilized to map items, item content features, and behavior types into embedding vectors:
\begin{equation}
p_n=W^I i_n,\quad q_n=W^C c_n,\quad r_n= W^B b_n, \label{emb1}
\end{equation}
where $W$ matrices are embedding matrices. $i_n$ and $b_n$ are one-hot vectors. $c_n$ can be a one-hot or multi-hot vector. Since we only use item category information which is widely used in previous studies~\cite{tanjim2020attentive,zhou2018atrank}, $c_n$ is a one-hot vector in this paper. If other item content information was available, it could also be included. The time interval between $x_n$ and $x_{n+1}$ is calculated as $\Delta t_n=t_{n+1}-t_n$. Following previous studies~\cite{zhou2018atrank}, we divide the overall time intervals into buckets and map each one-hot bucket vector into an embedding vector $s_n$ with $W^\Delta$:
\begin{equation}
    s_n=W^\Delta \mathrm{bucketize}(\Delta t_n). \label{emb2}
\end{equation}

\subsection{Multi-aspect Preference Modeling}
Stable preferences are with fewer fluctuations and can be reflected by historical target behavior sequences. In the context of a multi-typed behavioral sequence $\{x_1, x_2, \cdots , x_N\}$ of length $N$, we pick up target behaviors to form a target behavior sequence $\{\bar{x}_1, \bar{x}_2, \cdots , \bar{x}_M\}$~\footnote{We use $\bar{x}_\star$ to represent a target behavior.}, $M<N$. Reviewing Figure~\ref{fig:illu}, the multi-typed behavioral sequence of User 1 can be represented as $\{x_1, x_2, x_3, x_4\}$. $x_1$ and $x_2$ are target behaviors and the target behavior sequence is represented as $\{\bar{x}_1, \bar{x}_2\}$. After the embedding layer, the item and item content embeddings are sent to an LSTM layer: 
\begin{equation}
    h^S_m=\mathrm{LSTM}([p_m,q_m],h^S_{m-1}), \quad 1\leq m\leq M, \label{lst}
\end{equation}
where $[,]$ represents the concatenation of vectors and $h^S_m$ is the hidden state at the $m$-th step. $p_m,q_m$ are calculated with Equation~\eqref{emb1}.

User preferences are multi-aspect. To model multi-aspect preferences, we propose a multi-aspect projection mechanism. More specifically, we project the hidden state $h^S_M$ to multiple semantic subspaces, i.e., aspects:
\begin{equation}
    \tilde{h}^S_{j}=\mathcal{P}_j( h^S_M ), \quad 1\leq j\leq J,
\end{equation}
where $\mathcal{P}_j$ is the projection function for the $j$-th implicit aspect and $\tilde{h}^S_{j}$ is the projected preference representation. The rationality of the multi-aspect projection mechanism is easy to be proved. Many studies~\cite{zeiler2014visualizing} have proved that different filters in a convolutional layer focus on distinct features of an image. Our projection functions are similar to filters. Each projection function focuses on automatically extracting characteristics of a specific aspect. Reviewing Figure~\ref{fig:illu}, we should project user preferences to two semantic subspaces: the category-aspect and the brand-aspect. Then, we obtain user preference representations of User 1: $\tilde{h}^S_{1}$ and $\tilde{h}^S_{2}$.

The projection functions can either be linear or nonlinear. For simplicity, we present the linear projection function:
\begin{equation}
    \tilde{h}^S_{j}=W^P_{j} h^S_M, \quad 1\leq j\leq J,
    \label{mal}
\end{equation}
where $W^P$ are the projection matrices.

\subsection{Multi-aspect Intent Modeling}
Dynamic purchase intents evolve over time and can be reflected by multi-typed behavioral sequences~\cite{tanjim2020attentive}. A common scenario is that when a user wants something, she will click some related items and then add some of these items to her cart. 

Behavior type and time interval are behavioral specifics which provide a fine-grained understanding of one user interaction on one item. Although behavior specifics cannot provide any information about the item, they reflect the importance of each interaction. To consider behavioral specifics while modeling an item sequence, we regard behavioral specifics as strong signals in input, forget, and output gates. That is to say, behavioral specifics are involved in controlling item sequence modeling. This variant of LSTM is called behavior-enhanced LSTM, which is formulated as follows:
\begin{align}
    i_{n}=&\sigma(W_{i}[p_{n},q_{n},r_{n},s_{n},h_{n-1}]+\hat{b}_{i}), \label{input}\\
    f_{n}=&\sigma(W_{f}[p_{n},q_{n},r_{n},s_{n},h_{n-1}]+\hat{b}_{f}), \label{forget}\\
    \hat{c}_{n}=&f_{n}\odot\hat{c}_{n-1}+i_{n}\odot \tanh(W_{\hat{c}}[p_{n},q_{n},h_{n-1}]+\hat{b}_{\hat{c}}), \label{onlyinput}\\
    o_{n}=&\sigma(W_{o}[p_{n},q_{n},r_{n},s_{n},h_{n-1}]+\hat{b}_{o}), \label{output}\\
    h_{n}=&o_{n}\odot \tanh(\hat{c}_{n}),\label{outputh}
\end{align}
where $i_{n}$, $f_{n}$ and $o_{n}$ are gates at the $n$-th step. $\hat{c}_{n}$ is the cell memory. $W$ matrices and $\hat{b}$ terms are trainable parameters. $\sigma$ is the element-wise sigmoid function. $\odot$ is the element-wise product. $p_{n}$ and $q_{n}$ are item characteristics and calculated with Equation~\eqref{emb1}. $r_{n}$ and $s_{n}$ are behavioral specifics and can be calculated with Equation~\eqref{emb1} and Equation~\eqref{emb2}, respectively.


\begin{figure}[t]
\centering
    \includegraphics[width= 7cm]{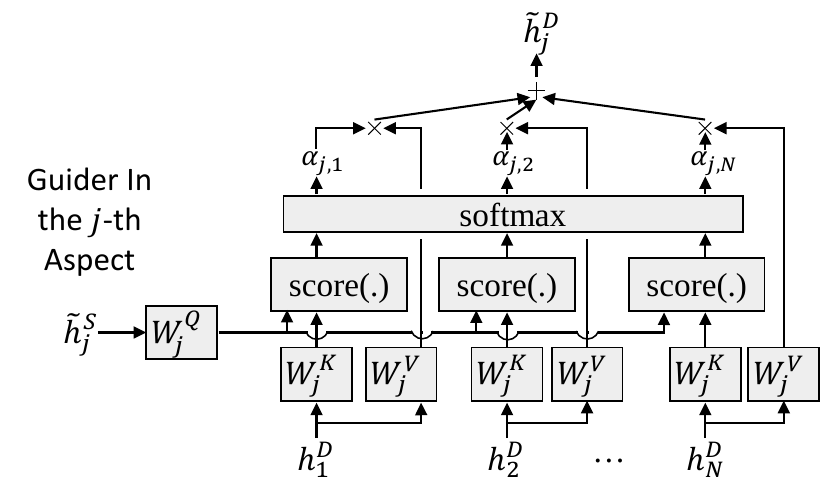}
    \caption{Refinement Attention Mechanism.}
    \label{fig:ratt}
\end{figure}

When modeling multi-typed behavioral sequences, it is important to consider the presence of noise. For example, click behaviors can be quite noisy due to accidental touches. To mitigate this issue, we regard stable preferences as guiders. In detail, we employ a refinement attention mechanism in each aspect. The attention mechanism utilizes the stable preference as a guider to guide the dynamic intent extraction from the sequence by assigning different weights. In this way, we will learn $J$ intent representations from $J$ aspects. Refinement attention mechanisms constitute the multi-aspect refinement attention mechanism, which is inspired by the success of the multi-head self-attention~\cite{vaswani2017attention}. Refinement attention mechanism is formulated as follows and the detailed structure is illustrated in Figure~\ref{fig:ratt}: 
\begin{align}
    \mathrm{score}(\tilde{h}^S_{j},h^D_{n}) =&v^\top_{j}\tanh(W^Q_{j}\tilde{h}^S_{j}+W^K_{j}h^D_{n}+\hat{b}^A_{j}), \label{score}\\
    \alpha_{j,n} =&\frac{\exp(\mathrm{score}(\tilde{h}^S_{j},h^D_{n}))}{\sum_{n^\prime =1}^N \exp(\mathrm{score}(\tilde{h}^S_{j},h^D_{n^\prime}))}, \label{alpha}\\
    \tilde{h}^D_{j} =&\sum_{n=1}^N \alpha_{j,n} W^V_{j}h^D_{n}, \label{mal1}
\end{align}
where $1\leq j\leq J$, $\tilde{h}^S_{j}$ is calculated with Equation~\eqref{mal}, $h^D_{n}$ is calculated with Equation~\eqref{outputh}, $W$ matrices denote weight matrices, $v_{j}$ is the weight vector, and $\hat{b}^A_{j}$ is the bias vector. $\tilde{h}^D_{j}$ is the intent representation in the $j$-th aspect. Reviewing Figure~\ref{fig:illu}, we can obtain user intent representations $\tilde{h}^D_{1}$ and $\tilde{h}^D_{2}$ with the help of $\tilde{h}^S_{1}$ and $\tilde{h}^S_{2}$, respectively.

\subsection{Preference and Intent Fusing}
The significance of preferences/intents is uncertain. We propose a multi-aspect gated fusion mechanism to balance the influence of preferences and intents. The mechanism considers characteristics of preferences and intents in each aspect and constructs the hybrid user representation $\tilde{h}^H_{j}$ accordingly:
\begin{align}
    \beta_j =& \sigma(w^\top_j[\tilde{h}^S_{j},\tilde{h}^D_{j}]+\hat{b}^F_j), \label{beta} \\
    \tilde{h}^H_{j} =& (1-\beta_j) \tilde{h}^S_{j} + \beta_j \tilde{h}^D_{j}, \label{hH}
\end{align}
where $1\leq j\leq J$, $w_j$ is the weight vector, and $\hat{b}^F_j$ is the bias. Reviewing Figure~\ref{fig:illu}, we can integrate $\tilde{h}^D_{1}$ and $\tilde{h}^S_{1}$, resulting in the generation of $\tilde{h}^H_{1}$. Similarly, by integrating $\tilde{h}^D_{2}$ and $\tilde{h}^S_{2}$, we can obtain $\tilde{h}^H_{2}$.

Transformer~\cite{vaswani2017attention} proposes leveraging a concatenation and a projection to integrate parallel heads. Similarly, we concatenate hybrid user representations in different aspects (i.e., semantic subspaces) and utilize a projection again. In this way, we obtain the final user representation $h^F$ as the following equation shows:
\begin{equation}
    h^F = W^\rho[\tilde{h}^H_{1},\tilde{h}^H_{2},\cdots,\tilde{h}^H_{J}], \label{hF}
\end{equation}
where the projection is a parameter matrix $W^\rho$. The dimensions of $h^F$ and $\tilde{h}^H_{\star}$ are equal. Reviewing Figure~\ref{fig:illu}, we can derive the final user representation $h^F$ of User 1 based on $\tilde{h}^H_{1}$ and $\tilde{h}^H_{2}$.

Inspired by studies~\cite{chen2019matching,LiCLYH21}, we predict next item and next category that a user is interested in with the following equations:
\begin{align}
    \hat{y}^I =& \mathrm{softmax}(W^{O,I} h^F), \label{yi} \\ 
    \hat{y}^C =& \mathrm{softmax}(W^{O,C} h^F), \label{yc}
\end{align}
where $W^{O,I}$ and $W^{O,C}$ are the weight matrices. $\hat{y}^I$ and $\hat{y}^C$ are the predicted probabilities of items and categories, respectively.

\subsection{Model Training}




To augment the training data, we predict next item $i_{m+1}$ and category $c_{m+1}$ at every time step $m$. Thus, the total loss function is given by the following equation:

\begin{equation}
\mathcal{L}(\Theta)=-\sum_{u\in U} \sum_{m=1}^M \bigg(y^I_{m+1}\log(\hat{y}^I_{m+1})+\gamma y^C_{m+1}\log(\hat{y}^C_{m+1})\bigg)+\lambda \left \| \Theta \right \|^2,
\end{equation}
where $\Theta$ denotes all trainable parameters, $\gamma$ is the hyper-parameter which controls the effectiveness of category information, and $\lambda$ is the hyper-parameter which controls the strength of the L2 regularization to prevent overfitting. Besides the L2 regularization, we adopt dropout~\cite{srivastava2014dropout} in the fusing component. 

We optimize all models with Adam optimizer~\cite{kingma2014adam}, which is an adaptive learning rate optimization algorithm. 


\section{Experiments}\label{ex}
Next, we answer the following research questions: 
\begin{itemize}
    \item \textbf{RQ1:} How does MAINT perform compared with various advanced recommender models?
    \item \textbf{RQ2:} Are the support types of behaviors helpful for improving the performance of predicting the next target behavior?
    \item \textbf{RQ3:} How do different designs contribute to the model performance?
    \item \textbf{RQ4:} How do hyper-parameters affect the performance?
    \item \textbf{RQ5:} How is the interpretation ability of our model?
\end{itemize}
\subsection{Experimental Settings}
\subsubsection{Datasets and Preprocessing}
We utilize two real-world e-commerce datasets, i.e., Taobao~\footnote{https://tianchi.aliyun.com/dataset/dataDetail?dataId=46} and Retailrocket~\footnote{https://www.kaggle.com/retailrocket/ecommerce-dataset}, which contain multiple behavior types. 
Dataset statistics are shown in Table~\ref{tab:stat}. Similar to previous studies~\cite{li2018learning,wang2019modeling}, we filter out users who have fewer than ten interactions and items that appear fewer than twenty times for Taobao. For Retailrocket, we filter out users and items with fewer than five and ten records, respectively. In addition, each user should have at least one buy record. Category information provided by Retailrocket is hierarchical (tree-like). In our experiments, we only use category information of the lowest level. The maximum length of sequences is twenty in our experiments.

\begin{table}[t]
\caption{Dataset statistics.}
\centering
\small
\label{tab:stat}
    \begin{tabular}{lrr}
    \toprule
    \textbf{Dataset} & \textbf{Taobao}                                                                                                          & \textbf{Retailrocket}                                                                             \\ \hline
    Users            & 10,000                                                                                                                   & 1,407,580                                                                                         \\
    Items           & 2,876,947                                                                                                                & 417,053                                                                                           \\
    Categories       & 8,916                                                                                                                    & 1,086                                                                                             \\ \cline{2-3} 
    Interactions     & \begin{tabular}[c]{@{}r@{}}11,550,581 \ \ \ Clicks\\ 242,556 \ Collects\\ 343,564 \ \ \ \ Carts\\ 120,205 \ \ \ \ \ Buys\end{tabular} & \begin{tabular}[c]{@{}r@{}}2,664,312 \ Views\\ 69,332 \ \ Carts\\ 22,457 \ \ \ Buys\end{tabular} \\ \cline{2-3}
    \# Behavior Types & 4 & 3 \\ \bottomrule
    \end{tabular}
\end{table}

\subsubsection{Evaluation Protocols}
Following the leave-one-out strategy applied widely on existing studies~\cite{he2017neurala,gao2021learning}, we utilize the most recently bought item of each user for testing, the second recently bought item for validation, and the rest bought items for training. For efficient evaluation, each positive instance is paired with 100 non-interactive items~\cite{tanjim2020attentive,he2017neurala}. A good recommender system should recommend new items that a user has not consumed before~\cite{gao2021learning}. So for each user, we delete her training records with the test item.

To evaluate the performance of each model, we use two metrics: Hit Ratio (HR@K) and Normalized Discounted Cumulative Gain (NDCG@K). K is the number of recommended items. There are two primary reasons for using these two metrics. First, these metrics have been widely used by previous researchers~\cite{he2017neurala,gao2021learning}, and employing the same metrics enables a fair comparison of models. Second, these metrics focus on different perspectives: NDCG is a ranking-based measure that focuses on the position of the test item in the Top-K recommendation list, while HR is a recall-based measure that determines whether the test item appears in the Top-K recommendation list. Together, they provide a comprehensive evaluation of a model from different perspectives.

\subsubsection{Methods for Comparison}
We compare our proposed method with three groups of recommender models: 

\textbf{Multi-behavioral Non-sequential Recommender Models.} 
\begin{itemize}
    \item \textbf{BF}~\cite{zhao2015improving}. This method jointly factorizes matrices of different behavior types with sharing item-side embeddings.
    \item \textbf{MC-BPR}~\cite{loni2016bayesian}. This method adopts sampling rules considering levels of different behavior types.
    \item \textbf{NMTR}~\cite{gao2021learning}. This method considers the cascading relationship among different behavior types. It shares user-side and item-side embeddings and learns a separate interaction function for each behavior type. 
\end{itemize}

\textbf{Single-behavioral Sequential Recommender Models.} 
\begin{itemize}
    \item \textbf{GRU4Rec}~\cite{hidasi2016session}. This method uses a GRU to model behaviors.
    \item \textbf{MCPRN}~\cite{wang2019modeling}. This method uses multiple RNNs to learn multiple purposes and proposes a purpose router to decide which RNN should be updated with an interacted item. Vanilla MCPRN does not consider category information. To ensure a fair comparison, we extend it by concatenating item embeddings and category embeddings as the input.
    \item \textbf{HGN}\cite{MaKL19}. This method designs a feature gating and an instance gating to select features and items, respectively. We extend vanilla HGN to consider category information, too.
\end{itemize}

\textbf{Multi-behavioral Sequential Recommender Models.} 
\begin{itemize}
    \item \textbf{ATRank}~\cite{zhou2018atrank}. This method groups behaviors by behavior type and chooses a self-attention network to model the influence among different behavior types.
    \item \textbf{BINN}~\cite{li2018learning}. This method exploits an LSTM-based model to learn from the preference behaviors and the session behaviors. We extend vanilla BINN to consider category information.
    \item \textbf{ASLI}~\cite{tanjim2020attentive}. This method learns item similarities based on the interacted item sequence via a transformer layer and obtains the user’s intent based on her actions on a particular category via a convolution layer.
    \item \textbf{IARS}~\cite{XuZY21}. This method leverages multiple GRUs to learn diverse intentions. It designs three tasks to improve recommendation performance.
\end{itemize}

HGN~\cite{MaKL19} and ASLI~\cite{tanjim2020attentive} have outperformed some sequential models (such as FPMC~\cite{rendle2010factorizing}, Fossil~\cite{HeM16}, Caser~\cite{DBLP:conf/wsdm/TangW18}, NextItNet~\cite{DBLP:conf/wsdm/YuanKAJ019}, SASRec~\cite{kang2018self}). We omit comparisons against these models.

\subsubsection{Implementation Details}
We implement MAINT with TensorFlow. The item embedding size is fixed to 64 for all models, which is suitable for models to learn a strong representation. The batch size of samples is 512. The initial learning rate is 0.01. $\lambda$ is set to $10^{-5}$. The number of implicit aspects is 3 (i.e., $J=3$). $\gamma$ is $1$. The dropout rate is 0.2. For all models, we tune their hyper-parameters and report the best performance.

\begin{table*}[!t]
\caption{Comparisons on Taobao and Retailrocket. $^\star$ means significantly better than the best baseline ($p<0.05$).}
\label{tab:re}
\centering
\small
    \begin{tabular}{l|rrrrrr}
    \hline
    \multirow{3}{*}{Method} & \multicolumn{6}{c}{Taobao}                                                                                                                          \\ \cline{2-7} 
                            & \multicolumn{2}{c|}{K=2}                               & \multicolumn{2}{c|}{K=6}                               & \multicolumn{2}{c}{K=10}          \\ \cline{2-7} 
                            & HR              & \multicolumn{1}{r|}{NDCG}            & HR              & \multicolumn{1}{r|}{NDCG}            & HR              & NDCG            \\ \hline
    BF                      & 0.1142          & \multicolumn{1}{r|}{0.1015}          & 0.2053          & \multicolumn{1}{r|}{0.1399}          & 0.2670          & 0.1592          \\
    MC-BPR                  & 0.1201          & \multicolumn{1}{r|}{0.1009}          & 0.2421          & \multicolumn{1}{r|}{0.1532}          & 0.3264          & 0.1794          \\
    NMTR                    & 0.1224          & \multicolumn{1}{r|}{0.1073}          & 0.2502          & \multicolumn{1}{r|}{0.1613}          & 0.3327          & 0.1871          \\ \hline
    GRU4Rec                 & 0.0989          & \multicolumn{1}{r|}{0.0850}          & 0.2113          & \multicolumn{1}{r|}{0.1328}          & 0.2971          & 0.1600          \\
    MCPRN                   & 0.2753          & \multicolumn{1}{r|}{0.2564}          & 0.3968          & \multicolumn{1}{r|}{0.3114}          & 0.4621          & 0.3328          \\
    HGN                     & 0.2552          & \multicolumn{1}{r|}{0.2381}          & 0.3898          & \multicolumn{1}{r|}{0.2958}          & 0.4655          & 0.3210          \\ \hline
    ATRank                  & 0.2588          & \multicolumn{1}{r|}{0.2275}          & 0.3753          & \multicolumn{1}{r|}{0.2857}          & 0.4459          & 0.3111          \\
    BINN                    & 0.2871          & \multicolumn{1}{r|}{0.2658}          & 0.3945          & \multicolumn{1}{r|}{0.3272}          & 0.4599          & 0.3562          \\
    ASLI                    & 0.2186          & \multicolumn{1}{r|}{0.2019}          & 0.3626          & \multicolumn{1}{r|}{0.2754}          & 0.4424          & 0.3027          \\
    IARS-S                  & \textit{0.3020} & \multicolumn{1}{r|}{\textit{0.2740}} & \textit{0.4095} & \multicolumn{1}{r|}{\textit{0.3278}} & \textit{0.4857} & \textit{0.3568} \\    
    \textbf{MAINT}          & $\textbf{0.3078}^\star$ & \multicolumn{1}{r|}{$\textbf{0.2773}^\star$} & $\textbf{0.4285}^\star$ & \multicolumn{1}{r|}{$\textbf{0.3289}^\star$} & $\textbf{0.5130}^\star$ & $\textbf{0.3582}^\star$ \\ \hline\hline
    \multirow{3}{*}{Method} & \multicolumn{6}{c}{Retailrocket}                                                                                                                    \\ \cline{2-7} 
                            & \multicolumn{2}{c|}{K=2}                               & \multicolumn{2}{c|}{K=6}                               & \multicolumn{2}{c}{K=10}          \\ \cline{2-7} 
                            & HR              & \multicolumn{1}{r|}{NDCG}            & HR              & \multicolumn{1}{r|}{NDCG}            & HR              & NDCG            \\ \hline
    BF                      & 0.1039          & \multicolumn{1}{r|}{0.1028}          & 0.2174          & \multicolumn{1}{r|}{0.1363}          & 0.3104          & 0.1654          \\
    MC-BPR                  & 0.1953          & \multicolumn{1}{r|}{0.1710}          & 0.3396          & \multicolumn{1}{r|}{0.2337}          & 0.4193          & 0.2587          \\
    NMTR                    & 0.2680          & \multicolumn{1}{r|}{0.2387}          & 0.4166          & \multicolumn{1}{r|}{0.3022}          & 0.5041          & 0.3295          \\ \hline
    GRU4Rec                 & 0.1030          & \multicolumn{1}{r|}{0.0965}          & 0.1827          & \multicolumn{1}{r|}{0.1245}          & 0.2323          & 0.1436          \\
    MCPRN                   & 0.3071          & \multicolumn{1}{r|}{0.2944}          & 0.4969          & \multicolumn{1}{r|}{0.3828}          & 0.6106          & 0.4207          \\
    HGN                     & 0.3827          & \multicolumn{1}{r|}{0.3461}          & 0.5574          & \multicolumn{1}{r|}{0.4215}          & 0.6305          & 0.4443          \\ \hline
    ATRank                  & 0.3686          & \multicolumn{1}{r|}{0.3573}          & 0.5363          & \multicolumn{1}{r|}{0.4246}          & 0.6413          & 0.4525          \\
    BINN                    & 0.3564          & \multicolumn{1}{r|}{0.3418}          & 0.5228          & \multicolumn{1}{r|}{0.4151}          & 0.6231          & 0.4370          \\
    ASLI                    & 0.4165          & \multicolumn{1}{r|}{0.3862}          & 0.5527          & \multicolumn{1}{r|}{0.4460}          & 0.6423          & 0.4698          \\
    IARS-S                  & \textit{0.4179} & \multicolumn{1}{r|}{\textit{0.3875}} & \textit{0.5634} & \multicolumn{1}{r|}{\textit{0.4482}} & \textit{0.6444} & \textit{0.4735} \\
    \textbf{MAINT}          & $\textbf{0.5187}^\star$ & \multicolumn{1}{r|}{$\textbf{0.4802}^\star$} & $\textbf{0.6525}^\star$ & \multicolumn{1}{r|}{$\textbf{0.5450}^\star$} & $\textbf{0.7175}^\star$ & $\textbf{0.5693}^\star$ \\ \hline
    \end{tabular}
\end{table*}

\subsection{Performance Comparison (RQ1)}
To answer \textbf{RQ1}, we repeat each experiment 5 times by changing the random seeds, and average results are reported in Table ~\ref{tab:re}. The two-tailed unpaired $t$-test is performed to detect significant differences between MAINT and the best baseline. We have the following findings:
\begin{itemize}
    \item In the first group, NMTR outperforms BF and MC-BPR due to its ability to model complex nonlinear interactions between users and items with deep learning.
    \item In the second group, MCPRN performs better than GRU4Rec. Since MCPRN divides items into several subsets according to user latent purposes and models each subset of items with a variant GRU. It could capture subset-level dependencies besides item-level dependencies. Compared with NMTR, although it ignores the behavioral diversity, it captures the sequence-related patterns of the item- and subset-level well and performs better. HGN also performs better than GRU4Rec and NMTR. HGN incorporates a feature gating, an instance gating, and an aggregation layer to capture feature-level, instance-level, and group-level relations between past and future interaction items. It also utilizes an item-item product module to capture item-item relations. It has better performance than MCPRN on Retailrocket. The reason may be that gating networks are good at processing sparse data.
    \item IARS performs best among baselines since it could model multiple coexisting intentions with multiple GRUs and attention layers. Besides, auxiliary tasks are helpful for solving the main task. BINN could learn users' historical preferences and present motivations considering both multi-behavioral and sequence-related patterns. ASLI could learn users' intents from their actions on a particular category. ATRank and ASLI have better performance than BINN on Retailrocket. The reason may be that ATRank and ASLI have attention-based units which are better at handling sparse data.
    \item MAINT achieves significant improvement over all baselines on all datasets in both HR and NDCG. Compared with MCPRN and ASLI, MAINT mainly can capture more aspects of user preferences/intents. Compared with HGN and BINN, MAINT excels in its ability to adaptively fuse preferences and intents while learning them from multiple aspects. Compared with ATRank, MAINT effectively models sequence-related patterns existing in the multi-typed behavioral sequence. Compared with IARS, MAINT can learn multi-aspect preferences and filter out noises existing in the multi-typed behavioral sequence.
\end{itemize}

\begin{figure}[t]
\centering
    \subfigure[Taobao]{
        \includegraphics[width= 4.5cm]{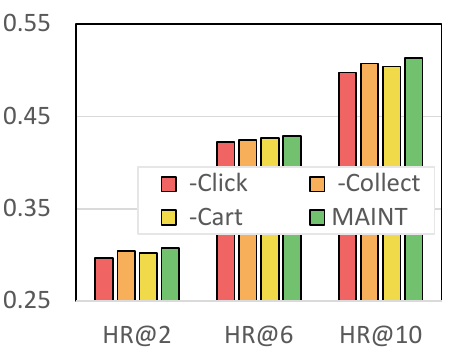}}
    \subfigure[Retailrocket]{
        \includegraphics[width= 4.5cm]{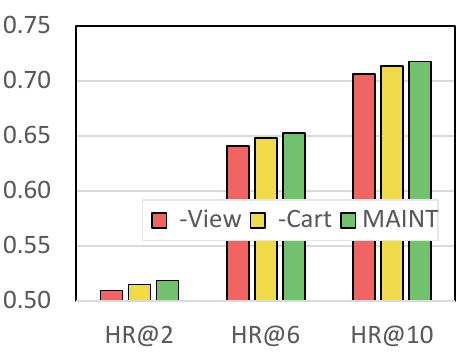}}
\caption{Impact study of different support types.}
\label{fig:sb}
\end{figure}

\subsection{Impact of support types of behaviors (RQ2)}
To answer \textbf{RQ2}, we design variants of our model. These variants are generated as follows: ``-'' behavior type means removing behaviors of this type. From the results in Figure~\ref{fig:sb}, we can see MAINT with all support types of behaviors achieves the best performance, which means support types of behaviors are helpful for improving prediction performance. Among support types, \emph{click} and \emph{view} are the most important behavior types to help MAINT improve performance. That is probably because behaviors of \emph{click} and \emph{view} account for a significant proportion.

\subsection{Ablation Study (RQ3)}
\begin{table}[b]
\caption{Ablation study of key designs in MAINT.}
\label{as}
\centering
\small
    \begin{tabular}{l|rr|rr}
    \toprule
    \multirow{2}{*}{Model} & \multicolumn{2}{c|}{Taobao} & \multicolumn{2}{c}{Retailrocket}   \\ \cline{2-5} 
                           & HR@10        & NDCG@10      & HR@10            & NDCG@10         \\ \hline
    MAINT-MP               & 0.4977       & 0.3471       & 0.7144           & 0.5560          \\
    MAINT-BLSTM            & 0.4977       & 0.3465       & 0.7066           & 0.5600          \\
    MAINT-RAtt             & 0.4876       & 0.3482       & 0.7109           & 0.5550          \\
    MAINT-MGFus            & 0.4925       & 0.3470       & 0.6938           & 0.5302          \\
    MAINT                  & 0.5130       & 0.3582       & 0.7175           & 0.5693          \\ \bottomrule
    \end{tabular}
\end{table}
To prove the effectiveness of the key designs in MAINT and answer \textbf{RQ3}, we consider different model variants from four perspectives:
\begin{itemize}
    \item Effect of the Multi-aspect Projection Mechanism. MAINT-MP. We consider one variant of MAINT without the projection mechanism. In other words, we utilize one latent vector to represent a user's preferences/intents (i.e., $J=1$).
    \item Effect of Behavior-enhanced LSTM. MAINT-BLSTM. We do an ablation study to test the effectiveness of Behavior-enhanced LSTM by comparing the performance of the model with vanilla LSTM.
    \item Effect of Refinement Attention Mechanism. MAINT-RAtt. We replace Refinement Attention Mechanism with vanilla attention~\cite{yang2016hierarchical}.
    \item Effect of Multi-aspect Gated Fusion Mechanism. MAINT-MGFus. We replace Multi-aspect Gated Fusion Mechanism with concatenation. 
\end{itemize}

Table~\ref{as} reports the results. We can find the full version of MAINT achieves the best performance in all cases. This proves that the key designs work. The reasons why our designs work have been introduced in the previous section.

\begin{figure}[b]
\centering
    \subfigure[Taobao.]{
        \includegraphics[width= 5cm]{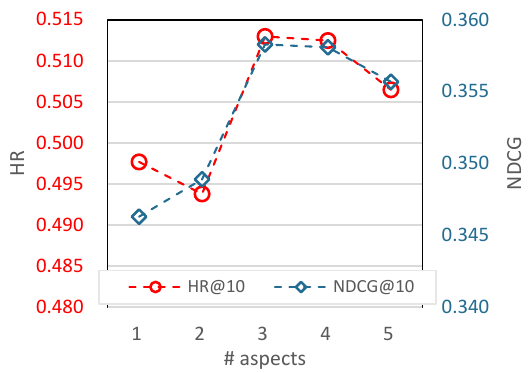}}
    \subfigure[Retailrocket.]{
        \includegraphics[width= 5cm]{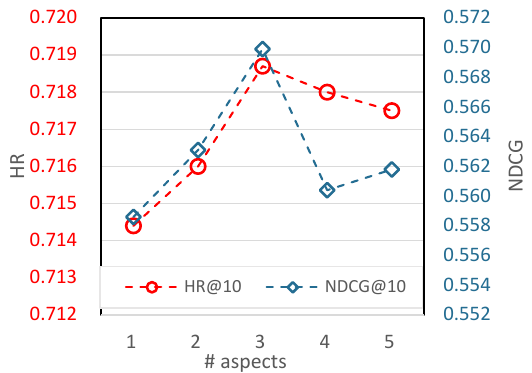}}    
\caption{Study of \# aspects.}
\label{fig:nss}
\end{figure}
\begin{figure}[t]
\centering
    \subfigure[Taobao.]{
        \includegraphics[width= 5cm]{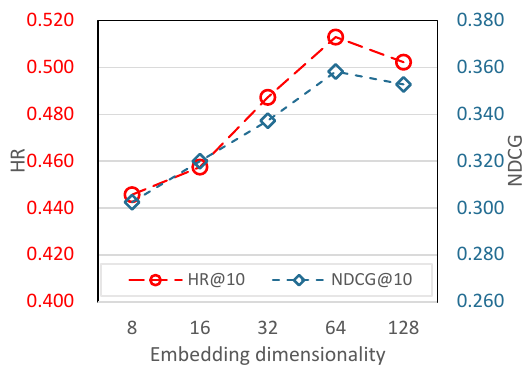}}
    \subfigure[Retailrocket.]{
        \includegraphics[width= 5cm]{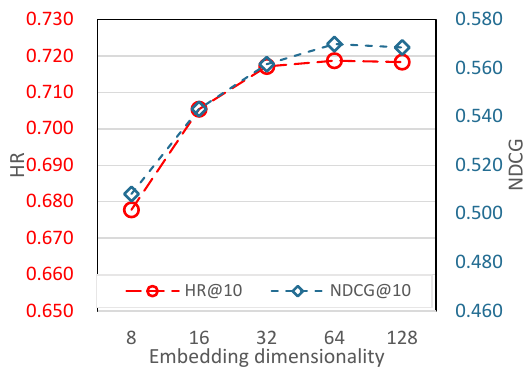}}        
\caption{Study of embedding dimensionality.}
\label{fig:ed}
\end{figure}
\begin{figure}[t]
\centering
    \subfigure[Taobao.]{
        \includegraphics[width= 5cm]{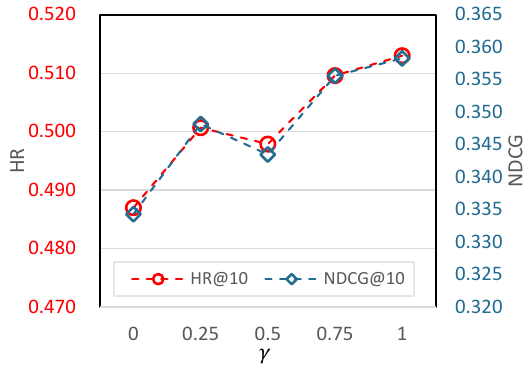}}
    \subfigure[Retailrocket.]{
        \includegraphics[width= 5cm]{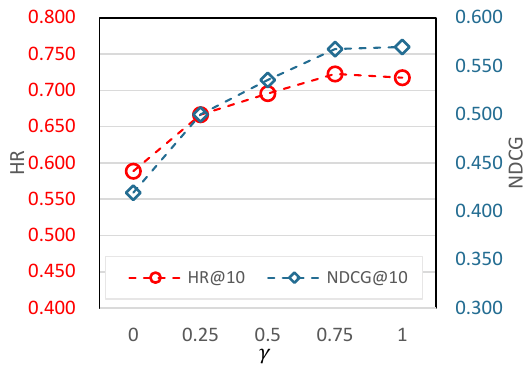}}
\caption{Study of $\gamma$.}
\label{fig:ga}
\end{figure}
\begin{figure}[t]
\centering
    \subfigure[Taobao.]{
        \includegraphics[width= 5cm]{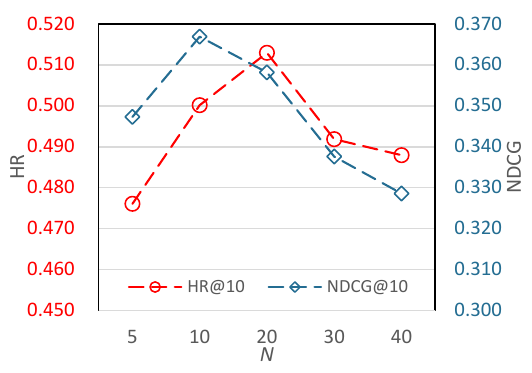}}
    \subfigure[Retailrocket.]{
        \includegraphics[width= 5cm]{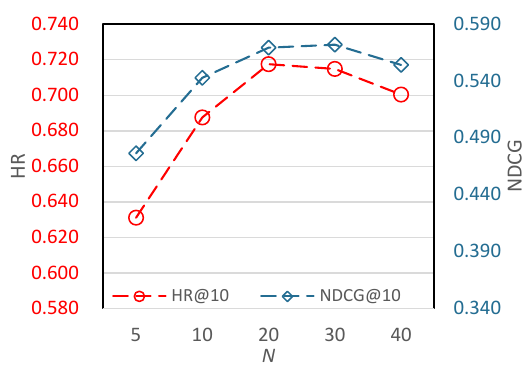}}
\caption{Study of the maximum length of sequences.}
\label{fig:ml}
\end{figure}

\subsection{Impact of Hyper-parameters (RQ4)}
For answering \textbf{RQ4}, we select four important hyper-parameters to study: 
\begin{itemize}
    \item Number of implicit aspects $J$. We adjust the number of aspects and the results are shown in Figure~\ref{fig:nss}. As the number of aspects increases, the performance grows in NDCG. When the number of aspects is 4, the performance drops. We think the overfitting problem may occur.
    \item Embedding dimensionality. We increase embedding dimensionality from 8 to 64, the model achieves better performance as Figure~\ref{fig:ed} shows. Because of the overfitting phenomenon, the performance degrades a little in NDCG with the further increase of embedding dimensionality.
    \item Coefficient $\gamma$. Figure~\ref{fig:ga} shows the results. Next-category recommendation task plays an important role in enhancing the performance of next-item recommendations. In general, the model performs better with larger values of $\gamma$. The reason may be that next-category recommendation task helps the model learn user preferences/intents from the category-aspect.
    \item Maximum length of sequences $N$. We conduct experiments by varying the sequence length between 5 and 40, and the results are presented in Figure~\ref{fig:ml}. We observed that as the length increases, the performance of MAINT initially improves but then declines. When the length is 10, MAINT achieves the best performance in NDCG on Taobao. When the length is 30, MAINT achieves the best performance in NDCG on Retailrocket. When the length is 20, MAINT achieves the best performance in HR on both datasets. It indicates that a longer sequence may include too many irrelevant items for predicting future items. In other words, the preference/intent drift problem is serious in a longer sequence. A larger length might lead to heavy computational costs at the same time. Meanwhile, the average lengths of processed sequences are about 19 and 17 on Taobao and Retailrocket, respectively. So the maximum length of sequences is 20 in our experiments.
\end{itemize}

\subsection{Case Study of model’s Interpretability (RQ5)}
\begin{figure}[t]
\centering
    \includegraphics[width= 12cm]{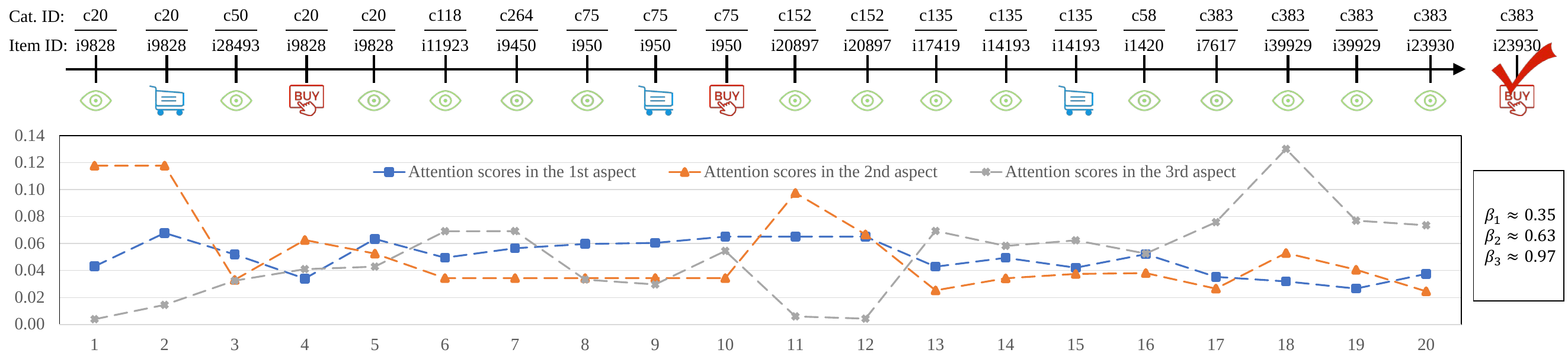}
    \caption{A real example from a sampled user on Retailrocket.}
    \label{fig:visualcase}
\end{figure}

In response to \textbf{RQ5}, we conduct a case study by randomly selecting a user from the Retailrocket dataset. The attention scores $\alpha$ and gate scores $\beta$ are visualized in Figure~\ref{fig:visualcase}. In this case, our model successfully recommends both the item and the category. Based on 20 behaviors of the user, our model initially recommends item i23930, which the user subsequently purchases. Our observations are as follows: 
\begin{itemize}
    \item MAINT focuses on different behaviors in different aspects since the distributions vary a lot in 3 aspects. In the 3rd aspect, the attention scores of the 17th, 18th, 19th, and 20th behaviors are high, and the categories of these items align with the predicted item. This suggests that the 3rd aspect may represent the category-aspect. However, due to insufficient information, it is not possible to infer the specific meanings represented by the other aspects.
    \item MAINT is capable of balancing preferences and intents, as indicated by the variation in gate scores across 3 aspects. In the 3rd aspect (i.e., the category-aspect), the gate score is high. This could be because the user made few purchases a long time ago but clicked many items recently, indicating that her intents may be more significant.
\end{itemize}

The experiment demonstrates the effectiveness of both the attention mechanism and the gated fusion mechanism in interpreting recommendation results.

\section{Discussion}\label{dis}
In this section, we discuss the extensibility and efficiency of our model.
\subsection{Model Extensibility}
\begin{figure}[t]
\centering
    \subfigure[]{
        \includegraphics[width= 4.5cm]{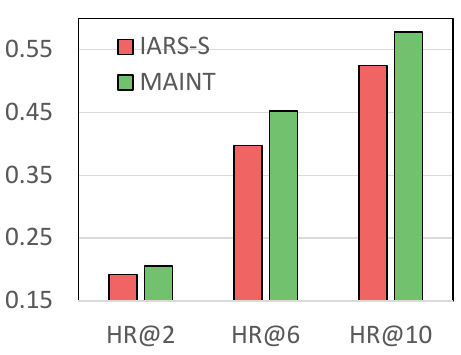}}
    \subfigure[]{
        \includegraphics[width= 4.5cm]{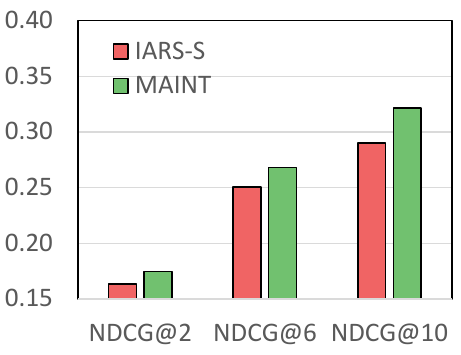}}
\caption{Comparisons on Movielens-1M.}
\label{fig:movierec}
\end{figure}

We would like to emphasize that our proposed model serves as a general recommender model, given its ability to model multi-typed behaviors. To show the extensibility of our model, we utilize another dataset: Movielens-1M~\footnote{https://grouplens.org/datasets/movielens/1m/}.  

Movielens-1M is a popular dataset~\cite{HarperK16,kang2018self}. It consists of 1,000,209 ratings for approximately 3,900 movies, provided by 6,040 users. The ratings are categorized into five levels (1, 2, 3, 4, 5), which can be viewed as five types of behavior. Our goal with this dataset is to predict the next movie that a user will rate 5 stars. For movies with multiple genre (i.e., category) tags, we only use the most frequently occurring one in the dataset. We also limit the sequence length to a maximum of 20.

We found that the model achieved the best results under the existing hyper-parameter settings. To provide a comparative analysis, we selected the most competitive baseline (i.e., IARS-S). The results are depicted in Figure~\ref{fig:movierec}, clearly demonstrating that our model continues to perform exceptionally well.

\subsection{Model Efficiency}
\begin{figure}[t]
\centering
    \subfigure[Offline Training.]{
        \includegraphics[width= 5cm]{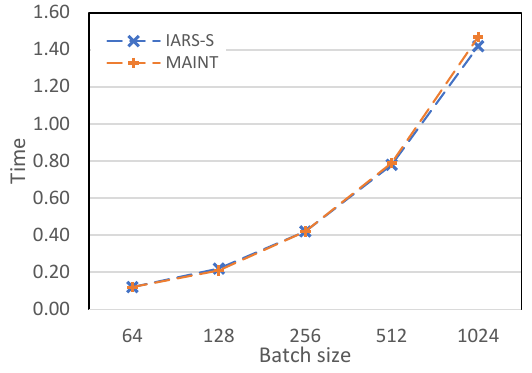}}
    \subfigure[Online Recommending.]{
        \includegraphics[width= 5cm]{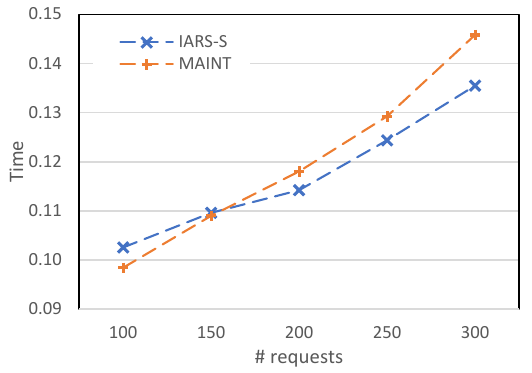}}
\caption{Time Comparison.}
\label{fig:timea}
\end{figure}

We then analyze the time complexity of our model. The time complexity of the shared embedding layer is $O(N)$. The time complexity of the multi-aspect projection mechanism is $O(Jd^2)$, where $d$ is the embedding dimensionality. The time complexity of two LSTMs is $O(Md^2+Nd^2)$, where $M<N$. The time complexity of the multi-aspect refinement attention mechanism is $O(JNd^2)$. The time complexity of the multi-aspect gated fusion mechanism is $O(Jd+Jd^2)$. Therefore, the total time complexity of our model is $O(N+Jd^2+Md^2+Nd^2+JNd^2+Jd+Jd^2)$, i.e., $O(JNd^2)$. The total time complexity of IARS-S is $O(\Lambda Nd^2)$, where $\Lambda$ is the number of GRUs. Figure~\ref{fig:timea} (a) compares the training time per batch of the best models (IARS-S and MAINT) in seconds. Figure~\ref{fig:timea} (b) compares the performance of these models in an online recommendation scenario, where the number of requests varies. We can see that the efficiency of MAINT is slightly worse than IARS-S. But the recommendation performance of MAINT is markedly superior. This limitation can be mitigated by utilizing machines with enhanced computational capabilities. On balance, the benefits of our model significantly outweigh its limitations.

We also analyze the space complexity of our model. The space complexity of the shared embedding layer is $O(\check{d}d)$, where $\check{d}$ is the sum of the dimensions of one-hot vectors. The space complexity of the multi-aspect projection mechanism is $O(Jd^2)$. The space complexity of two LSTMs is $O(d^2)$. The space complexity of the multi-aspect refinement attention mechanism is $O(Jd^2)$. The space complexity of the multi-aspect gated fusion mechanism is $O(Jd+Jd^2)$. Therefore, the space complexity of our model is $O(\check{d}d+Jd^2)$. The total space complexity of IARS-S is $O(\check{d}d+\Lambda d^2)$.

\section{Conclusion}\label{co}
In this paper, we propose a novel model called MAINT for multi-behavioral sequential recommendation. The experiments demonstrate that our models consistently outperform state-of-the-art recommender models. In summary, the key contributions are as follows: we propose an approach to extract multi-aspect preferences from target behaviors; we design network structures to extract multi-aspect intents from multi-typed behaviors; and we develop a mechanism to adaptively fuse multi-aspect preferences and intents. Pre-training has proven effective in various fields. Therefore, we aim to design an item embedding pre-training method (based on graph neural networks) considering multi-typed behaviors in the future.

\section*{Acknowledgements}
This research is supported in part by the National Key Research and Development Program of China (Grant No. 2020YFB1707701), the Natural Science Foundation of Shandong Province (ZR2023QF100, ZR2022QF050), the General Project for Undergraduate Education and Teaching Research at OUC (2023JY016), the National Science Foundation of China (No. 62072304, No. 62172277, No. 62172275 ), Shanghai Municipal Science and Technology Commission (No. 21511104700, No. 19511120300), the Shanghai East Talents Program, and the Oceanic Interdisciplinary Program of Shanghai Jiao Tong University (No. SL2020MS032).





\clearpage
\appendix
\appendixpage
\noindent\textbf{Details on the Conversion Rate.} The conversion rate of some behavior type is computed as:
\begin{equation*}
    Conversion\ Rate = \frac{\#\ behaviors\ of\ this\ type\ on\ each\ item\ before\ purchase}{\#\ behaviors\ of\ this\ type}.
\end{equation*}

\noindent\textbf{Training Procedure of MAINT.} The overall training procedure of MAINT is illustrated in Algorithm~\ref{trainalg}. Line 6-9 correspond to the multi-aspect preference modeling stage. Line 10-14 correspond to the multi-aspect intent modeling stage. Line 15-19 correspond to the preference and intent fusing stage.  
\begin{algorithm}[!t]
	\caption{Learning Algorithm for MAINT}
	\label{trainalg}
	\scriptsize
	\begin{algorithmic}[1] 
    \REQUIRE ~~\\ 
        multi-typed behavioral sequence of all users $\{X^u = \{x_1, x_2, \cdots , x_N\}\},u\in U$; item embedding size; number of implicit aspects $J$; category prediction loss coefficient $\gamma$; L2 regularization coefficient $\lambda$; dropout rate\\
    \ENSURE ~~\\ 
        MAINT with learned parameters $\Theta$
    \STATE Initialize all trainable parameters $\Theta$;
    \WHILE{stopping criteria is not met}
        \STATE Draw a mini-batch from $\{X^u\}$;
        \FOR{each $u$}
            \STATE $p_n,q_n,r_n,s_n=$ Embedding$(i_n,c_n,b_n,\Delta t_n), \quad 1\leq n\leq N$ (see Eq.~\eqref{emb1},~\eqref{emb2});
            \STATE $h^S_m=\mathrm{LSTM}([p_m,q_m],h^S_{m-1})$, the type of $\bar{x}_m$ is target behavior (see Eq.~\eqref{lst});
            \FOR{each $j \in [1,J]$}
                \STATE $\tilde{h}^S_{j}=W^P_{j} h^S_M$ (see Eq.~\eqref{mal});
            \ENDFOR
            \STATE $h^D_n=$ Behavior-enhanced\_LSTM($p_n,q_n,r_n,s_n$), $1\leq n\leq N$ (see Eq.~\eqref{input},\eqref{forget},\eqref{onlyinput},\eqref{output},\eqref{outputh});\
            \FOR{each $j \in [1,J]$}
                \STATE $\alpha_{j,n}=$ Refinement\_Attention\_Mechanism($\tilde{h}^S_{j},h^D_n$), $\quad 1\leq n\leq N$ (see Eq.~\eqref{alpha});\
                \STATE $\tilde{h}^D_j=\sum_{n=1}^N \alpha_{j,n} W^V_{j}h^D_{n}$ (see Eq.~\eqref{mal1});
            \ENDFOR
            \FOR{each $j \in [1,J]$}
                \STATE $\beta_j = \sigma(w^\top_j[\tilde{h}^S_{j},\tilde{h}^D_{j}]+\hat{b}^F_j)$ (see Eq.~\eqref{beta});
                \STATE $\tilde{h}^H_{j} = (1-\beta_j) \tilde{h}^S_{j} + \beta_j \tilde{h}^D_{j}$ (see Eq.~\eqref{hH});
            \ENDFOR
            \STATE $h^F = W^\rho[\tilde{h}^H_{1},\tilde{h}^H_{2},\cdots,\tilde{h}^H_{J}]$ (see Eq.~\eqref{hF});
            \STATE $\hat{y}^I = \mathrm{softmax}(W^{O,I} h^F), \quad \hat{y}^C = \mathrm{softmax}(W^{O,C} h^F)$ (see Eq.~\eqref{yi},~\eqref{yc});
        \ENDFOR
        \STATE Compute loss according to $\mathcal{L}(\Theta)=-\sum_{u\in U_{batch}} \sum_{m=1}^M \bigg(y^I_{m+1}\log(\hat{y}^I_{m+1})+\gamma y^C_{m+1}\log(\hat{y}^C_{m+1})\bigg)+\lambda \left \| \Theta \right \|^2$;
        \STATE Update $\Theta$ according to $\mathcal{L}$ and optimizer;
    \ENDWHILE
    \RETURN $\Theta$
	\end{algorithmic}
\end{algorithm} 


\end{document}